\documentclass[journal,twocolumn]{IEEEtran}

\usepackage[pdftex]{graphicx}
\usepackage{amsmath}
\usepackage{multirow}
\newtheorem{example}{\normalfont \textbf{Example}}
\newtheorem{lemma}{\normalfont \textbf{Lemma}}
\newtheorem{theorem}{\normalfont \textbf{Theorem}}

\newcommand{\bfb}{\mbox{\boldmath $b$}}
\newcommand{\bfc}{\mbox{\boldmath $c$}}
\newcommand{\bfd}{\mbox{\boldmath $d$}}

\newcommand{\bfg}{\mbox{\boldmath $g$}}
\newcommand{\bfx}{\mbox{\boldmath $x$}}
\newcommand{\bfy}{\mbox{\boldmath $y$}}

\begin{document}

\title{A Construction for Balancing Non-Binary Sequences Based on Gray Code Prefixes}

\author{Elie N. Mambou and Theo~G.~Swart,~\IEEEmembership{Senior Member,~IEEE}%
\thanks{This paper was presented in part at the IEEE International Symposium on Information Theory, Barcelona, Spain, July, 2016.}%
\thanks{The authors are with the Department of Electrical and Electronic Engineering Science, University of Johannesburg, P.~O.~Box 524, Auckland Park, 2006, South Africa (e-mails: \{emambou, tgswart\}@uj.ac.za).}
\thanks{This work is based on research supported in part by the National Research Foundation of South Africa (UID 77596).}}

\maketitle

\begin{abstract}
We introduce a new construction for the balancing of non-binary sequences that make use of Gray codes for prefix coding. Our construction provides full encoding and decoding of sequences, including the prefix. This construction is based on a generalization of Knuth's parallel balancing approach, which can handle very long information sequences. However, the overall sequence---composed of the information sequence, together with the prefix---must be balanced. This is reminiscent of Knuth's serial algorithm. The encoding of our construction does not make use of lookup tables, while the decoding process is simple and can be done in parallel. 
\end{abstract}


\IEEEpeerreviewmaketitle

\section{Introduction}\label{sec:1}

\IEEEPARstart{T}{he} use of balanced codes is crucial for some information transmission systems. Errors can occur in the process of storing data onto optical devices due to the low frequency of operation between structures of the servo and the data written on the disc. This can be avoided by using encoded balanced codes, as no low frequencies are observed. In such systems, balanced codes are also useful for tracking the data on the disc. Balanced codes are also used for countering cut-off at low frequencies in digital transmission through capacitive coupling or transformers. This cut-off is caused by multiple same-charge bits, and results in a DC level that charges the capacitor in the AC coupler \cite{immink2004}. In general, the suppression of low-frequency spectrum can be done with balanced codes.

A large body of work on balanced codes is derived from the simple algorithm for balancing sequences proposed by Knuth \cite{knuth1986}. According to Knuth's parallel algorithm, a binary sequence, $\bfx$, of even length $k$, can always be balanced by complementing its first or last $i$ bits, where $0 \leq i \leq k$. The index $i$ is then encoded as a balanced prefix that is appended to the data. The decoder can easily recover $i$ from the prefix, and then again complementing the first or last $i$ bits to obtain the original information. For Knuth's serial (or sequential) algorithm, the prefix is used to provide information regarding the information sequence's initial weight. Bits are sequentially complemented from one side of the overall sequence, until the information sequence and prefix together are balanced. Since the original weight is indicated by the prefix, the decoder simply has to sequentially complement the bits until this weight is attained.

Al-Bassam \cite{bassam1990phd} presented a generalization of Knuth's algorithm for binary codes, non-binary codes and semi-balanced codes (the latter occur where the number of 0's and 1's differs by at most a certain value in each sequence of the code). The balancing of binary codes with low DC level is based on DC-free coset codes. For the design of non-binary balanced codes, symbols in the information sequence are $q$-ary complemented from one side, but because this process does not guarantee balancing, an extra redundant symbol is added to enforce the balancing (similar to our approach later on). Information regarding how many symbols to complement is sent by using a balanced prefix.

Capocelli \emph{et al}. \cite{capocelli1991} proposed using two functions that must satisfy certain properties to encode any $q$-ary sequence into balanced sequences. The first function is similar to Knuth's serial scheme: it outputs a prefix sequence depending on the original sequence's weight. Additionally, all the $q$-ary sequences are partitioned into disjointed chains, where each chain's sequences have unique weights. The second function is then used to select an alternate sequence in the chain containing the original information sequence, such that the chosen prefix and the alternate sequence together are balanced.

Tallini and Vaccaro \cite{tallini1999} presented another construction for balanced $q$-ary sequences that makes use of balancing and compression. Sequences that are close to being balanced are encoded with a generalization of Knuth's serial scheme. Based on the weight of the information sequence, a prefix is chosen. Symbols are then ``complemented in stages'', one at a time, until the weight that balances the sequence and prefix together is attained. Other sequences are compressed with a uniquely decodable variable length code and balanced using the saved space.

Swart and Weber \cite{swart2009} extended Knuth's parallel balancing scheme to $q$-ary sequences with parallel decoding. However, this technique does not provide a prefix code implementation, with the assumption that small lookup tables can be used for this. Our approach aims to implement these prefixes via Gray codes. Swart and Weber's scheme will be expanded on in Section~\ref{sec:2.2}, as it also forms the basis of our proposed algorithm.

Swart and Immink \cite{swart2013} described a prefixless algorithm for balancing of $q$-ary sequences. By using the scheme from \cite{swart2009} and applying precoding to a very specific error correction code, it was shown that balancing can be achieved without the need for a prefix. 

Pelusi \emph{et al}. \cite{pelusi2015} presented a refined implementation of Knuth's algorithm for parallel decoding of $q$-ary balanced codes, similar to \cite{swart2009}. This method significantly improved \cite{capocelli1991} and \cite{tallini1999} in terms of complexity.

The rest of this paper is structured as follows. In Section~\ref{sec:2}, we present the background for our work, which includes Swart and Weber's balancing scheme for $q$-ary sequences \cite{swart2009} and non-binary Gray code theory \cite{guan1998}. In Section~\ref{sec:3}, a construction is presented for sequences where $k=q^t$. Section~\ref{sec:4} extends on our proposed construction to sequences with $k\neq q^t$. Finally, Section~\ref{sec:5} deals with the redundancy and complexity of our construction compared to prior art constructions, our conclusions are presented in Section~\ref{sec:6}.

\section{Preliminaries}\label{sec:2}

Let $\bfx = (x_1 x_2 \dots x_k)$ be a $q$-ary information sequence of length $k$, where $x_i \in \{0,1,\dots q-1\}$ is from a non-binary alphabet. A prefix of length $r$ is appended to $\bfx$. The prefix and information together are denoted by $\bfc = (c_1 c_2 \dots c_n)$ of length $n=k+r$, where $c_i \in \{0,1,\dots q-1\}$. Let $w(\bfc)$ refer to the weight of $\bfc$, that is the algebraic sum of symbols in $\bfc$. The sequence $\bfc$ is said to be balanced if
\begin{equation*}
    w(\bfc) = \sum_{i=1}^n c_i = \frac{n(q-1)}{2}.
\end{equation*}

Let $\beta_{n,q}$ represent this value obtained at the balancing state. For the rest of the paper, the parameters $k$, $n$, $q$ and $r$ are chosen in such a way that the balancing value, $\beta_{n,q}=n(q-1)/2$, leads to a positive integer.

\subsection{Balancing of $q$-ary Sequences}\label{sec:2.2}

Any information sequence, $\bfx$ of length $k$ and alphabet size $q$, can always be balanced by adding (modulo $q$) to that sequence one sequence from a set of balancing sequences \cite{swart2009}. The balancing sequence, $\bfb_{s,p} = (b_1 b_2 \dots b_k)$ is derived as
\begin{equation*}
    b_i = \begin{cases}
                s,           & i > p, \\
                s+1 \pmod q, & i \leq p,
          \end{cases}
\end{equation*}
where $s$ and $p$ are positive integers with $0 \leq s \leq q-1$ and $0 \leq p \leq k-1$. Let $z$ be the iterator through all possible balancing sequences, such that $z = sn + p$ and $0 \leq z \leq kq-1$. Let $\bfy$ refer to the resulting sequence when adding (modulo $q$) the balancing sequence to the information sequence, $\bfy = \bfx \oplus_q \bfb_{s,p}$, where $\oplus_q$ denotes modulo $q$ addition. The cardinality of balancing sequences equals $kq$ and amongst them, at least one leads to a balanced output $\bfy$.

Since $s$ and $p$ can easily be determined for the $z$-th balancing sequence using $z = sn + p$, we will use the simplified notation $\bfb_z$ to denote $\bfb_{s,p}$.\\

\begin{example}\label{ex1}
Let us consider the balancing of the 3-ary sequence 2101, of length 4. The encoding process is illustrated below, with weights in bold indicating that the sequences are balanced.
\begin{equation*}
	\begin{array}{c@{\quad\quad}c@{\;}c@{\;}c@{\;}c@{\;}c@{\quad\quad}c}
        z  & \bfx   & \oplus_q & \bfb_z & = & \bfy   & w(\bfy)    \\
        \hline
        0  & (2101) & \oplus_3 & (0000) & = & (2101) & \mathbf{4} \\
        1  & (2101) & \oplus_3 & (1000) & = & (0101) & 2          \\
        2  & (2101) & \oplus_3 & (1100) & = & (0201) & 3          \\
        3  & (2101) & \oplus_3 & (1110) & = & (0211) & \mathbf{4} \\
        4  & (2101) & \oplus_3 & (1111) & = & (0212) & 5          \\
        5  & (2101) & \oplus_3 & (2111) & = & (1212) & 6          \\
        6  & (2101) & \oplus_3 & (2211) & = & (1012) & \mathbf{4} \\
        7  & (2101) & \oplus_3 & (2221) & = & (1022) & 5          \\
        8  & (2101) & \oplus_3 & (2222) & = & (1020) & 3          \\
        9  & (2101) & \oplus_3 & (0222) & = & (2020) & \mathbf{4} \\
        10 & (2101) & \oplus_3 & (0022) & = & (2120) & 5          \\
        11 & (2101) & \oplus_3 & (0002) & = & (2100) & 3          \\
	\end{array}
\end{equation*}
For this example, there are four occurrences of balanced sequences.
\end{example}

\begin{figure}[!t]
    \centering
    \includegraphics{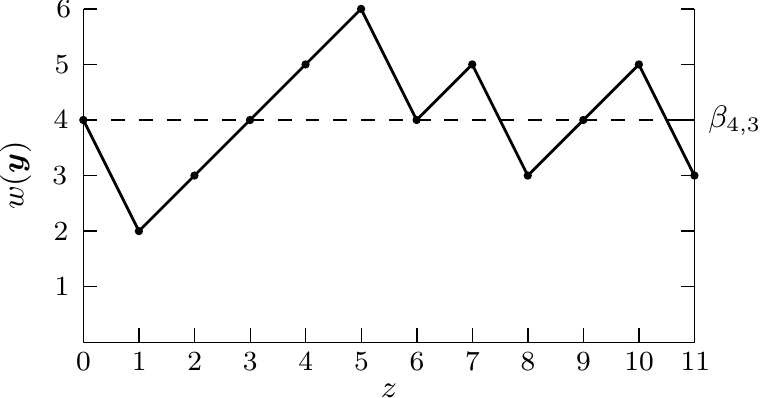}
    \caption{Random walk graph of $w(\bfy)$ for Example~\ref{ex1}}
    \label{fig:ex1}
\end{figure}

A $(\gamma, \tau)$-random walk refers to a path with random increases of $\gamma$ and decreases of $\tau$. In our case, a random walk graph is the plot of the function of $w(\bfy)$ versus $z$. In general, the random walk graph of $w(\bfy)$ always forms a $(1, q-1)$-random walk \cite{swart2009}. Fig.~\ref{fig:ex1} presents the $(1,2)$-random walk for Example~\ref{ex1}. The dashed line indicates the balancing value $\beta_{4,3} = 4$.
	
This method, as presented in \cite{swart2009}, assumed that the $z$ indices can be sent using balanced prefixes, but the actual encoding of these was not taken into account. For instance, in Example~\ref{ex1} indices $z=0$, $3$, $6$ and $9$ must be encoded into balanced prefixes, in order to send overall balanced sequences.

\subsection{Non-binary Gray Codes}\label{sec:2.3}

Binary Gray codes were first proposed by Gray \cite{gray1953} for solving problems in pulse code communication, and have been extended to various other applications. The assumption throughout this paper is that a Gray code is mapped from a set of possible sequences appearing in the normal lexicographical order. This ordering results in the main property of binary Gray codes: two adjacent codewords differ in only one bit.

The $(r',q)$-Gray code is a set of $q$-ary sequences of length $r'$ such that any two adjacent codewords differ in only one symbol position. This set is not unique, as any permutation of a symbol column within the code could also generate a new $(r',q)$-Gray code. In this work, a unique set of $(r',q)$-Gray codes is considered, as presented by Guan \cite{guan1998}. This set possesses an additional property: the difference between any two consecutive sequences' weights is $\pm1$. This same set of Gray codes was already determined in \cite{er1984} through a recursive method.

Let $\bfd = (d_1 d_2 \ldots d_{r'})$ be any sequence within the set of all $q$-ary sequences of length $r'$, listed in the normal lexicographic order. These sequences are mapped to $(r', q)$-Gray code sequences, $\bfg = (g_1 g_2 \ldots g_{r'})$, such that any two consecutive sequences are different in only one symbol position.

Table~\ref{gray} shows a $(3,3)$-Gray code, where $\bfd$ is the 3-ary representation of the index $z \in \{0,1,\ldots,26\}$ and $\bfg$ is the corresponding Gray code sequence. We see that for $\bfg$, the adjacent sequences' weights differ by $+1$ or $-1$.
\begin{table}[!b]
	\centering
	\caption{Example of $(3,3)$-Gray code}\label{gray}\vspace{-8pt}
	\begin{tabular}{ccc|ccc|ccc}
        \hline\hline
        $z$ & $\bfd$  & $\bfg$  & $z$ & $\bfd$  & $\bfg$  & $z$ & $\bfd$  & $\bfg$  \\
        \hline
        0   & $(000)$ & $(000)$ & 9   & $(100)$ & $(122)$ & 18  & $(200)$ & $(200)$ \\
        1   & $(001)$ & $(001)$ & 10  & $(101)$ & $(121)$ & 19  & $(201)$ & $(201)$ \\
        2   & $(002)$ & $(002)$ & 11  & $(102)$ & $(120)$ & 20  & $(202)$ & $(202)$ \\
        3   & $(010)$ & $(012)$ & 12  & $(110)$ & $(110)$ & 21  & $(210)$ & $(212)$ \\
        4   & $(011)$ & $(011)$ & 13  & $(111)$ & $(111)$ & 22  & $(211)$ & $(211)$ \\
        5   & $(012)$ & $(010)$ & 14  & $(112)$ & $(112)$ & 23  & $(212)$ & $(210)$ \\
        6   & $(020)$ & $(020)$ & 15  & $(120)$ & $(102)$ & 24  & $(220)$ & $(220)$ \\
        7   & $(021)$ & $(021)$ & 16  & $(121)$ & $(101)$ & 25  & $(221)$ & $(221)$ \\
        8   & $(022)$ & $(022)$ & 17  & $(122)$ & $(100)$ & 26  & $(222)$ & $(222)$ \\
        \hline\hline
	\end{tabular}	
\end{table}  

We will make use of the following encoding and decoding algorithms from \cite{guan1998}.

\subsubsection{Encoding algorithm for $(r',q)$-Gray code}

Let $\bfd = (d_1 d_2 \ldots d_{r'})$ and $\bfg = (g_1 g_2 \ldots g_{r'})$ denote respectively a $q$-ary sequence of length $r'$ and its corresponding Gray code sequence.

Let $S_i$ be the sum of the first $i-1$ symbols of $\bfg$, with $2\leq i \leq r'$ and $g_1=d_1$. Then
\begin{equation*}
    S_i=\sum_{j=1}^{i-1}g_j,
    \quad\text{and}\quad
    g_i = \begin{cases}
                d_i,     & \text{if } S_i \text{ is even}, \\
                q-1-d_i, & \text{if } S_i \text{ is odd}.
          \end{cases}
\end{equation*}
The parity of $S_i$ determines $\bfg$'s symbols from $\bfd$. If $S_i$ is even then the symbol stays the same, otherwise the $q$-ary complement of the symbol is taken.

\subsubsection{Decoding algorithm for $(r',q)$-Gray code}

Let $\bfg$, $\bfd$ and $S_i$ be defined as before, with $2\leq i \leq r'$ and $d_1=g_1$. Then
\begin{equation*}
    S_i=\sum_{j=1}^{i-1}g_j, \quad\text{and}\quad
    d_i = \begin{cases}
                g_i,     & \text{if } S_i \text{ is even}, \\
                q-1-g_i, & \text{if } S_i \text{ is odd}.
          \end{cases}
\end{equation*}

\section{Construction for $k=q^t$}\label{sec:3}

For the sake of simplicity, we will briefly explain the construction for information lengths limited to $k=q^t$, with $t$ being a positive integer. More details can be found in our conference paper \cite{mambou2016}. In the next section we will show how this restriction can be avoided.

The main component of this technique is to encode the balancing indices, $z$, into Gray code prefixes that can easily be encoded and decoded. The prefix together with the information sequence must be balanced. 

The condition, $k=q^t$, is enforced so that the cardinality of the $(r',q)$-Gray code is equal to that of the balancing sequences, making $r' = \log_q(kq) = \log_q(q^{t+1}) = t+1$.

\subsubsection{Encoding}

Let $\bfc' = (\bfg|\bfy) = (g_1 g_2 \ldots g_{r'} y_1 y_2 \ldots y_k)$ be the concatenation of the Gray code prefix with $\bfy$, with $|$ representing the concatenation. As stated earlier, for the sequences $\bfy$ we obtain a $(1,q-1)$-random walk, and for the Gray codes $\bfg$ we have a $(1,1)$-random walk. Therefore, when we concatenate the two sequences together, the random walk graph of $\bfc'$ forms a $(\{0;2\}, \{q-2;q\})$-random walk, i.e. increases of 0 or 2 and decreases of $q-2$ or $q$.

This concatenation of a Gray code prefix, $\bfg$, with an output sequence, $\bfy$, does not guarantee the balancing of the overall sequence, since the increases of 2 in the random walk graph do not guarantee that it will pass through a specific point. An extra symbol $u$ is added to ensure overall balancing, with $u = \beta_{n,q} - w(\bfc')$ if $0 \leq u \leq q-1$, otherwise $u=0$, thus forcing the random graph to a specific point. The overall sequence is the concatenation of $u$, $\bfg$ and $\bfy$, i.e. $\bfc = (u|\bfg|\bfy) = (u g_1 g_2 \ldots g_{r'} y_1 y_2 \ldots y_k)$. The length of $\bfc$ is $n=k+r'+1$.

In summary, the balancing of any $q$-ary sequence of length $k$, where $k=q^t$, can be achieved by adding (modulo $q$) an appropriate balancing sequence, $\bfb_z$, and prefixing a redundant symbol $u$ with a Gray code sequence, $\bfg$. The construction relies on finding a Gray code prefix to describe $z$, and at the same time be balanced together with $\bfy$.\\

\begin{figure}[!b]
	\centering
	\includegraphics{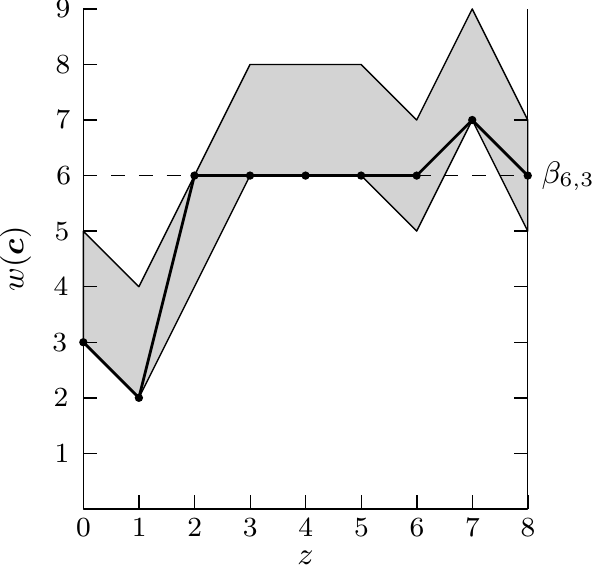}
	\caption{Random walk graph of $w(\bfc)$ for Example~\ref{ex2}}
	\label{fig:ex2}
\end{figure}

\begin{example}\label{ex2}
Let us consider the encoding of the ternary sequence, 201 of length 3. Since $t=1$, the length of Gray code prefixes will be $r'=2$. The overall length is $n=6$ and the balancing value is $\beta_{6,3}=6$. The encoding process below is followed.
\begin{equation*}
	\begin{array}{c@{\quad\quad}c@{\;}c@{\;}c@{\;}c@{\;}c@{\quad\quad}c@{\quad\quad}c}
        z & \bfx  & \oplus_q & \bfb_z & = & \bfy  & \bfc                           & w(\bfc)    \\
        \hline
        0 & (201) & \oplus_3 & (000)  & = & (201) & (\underline{\mathbf{0}00} 201) & 3          \\
        1 & (201) & \oplus_3 & (100)  & = & (001) & (\underline{\mathbf{0}01} 001) & 2          \\
        2 & (201) & \oplus_3 & (110)  & = & (011) & (\underline{\mathbf{2}02} 011) & \mathbf{6} \\
        3 & (201) & \oplus_3 & (111)  & = & (012) & (\underline{\mathbf{0}12} 012) & \mathbf{6} \\
        4 & (201) & \oplus_3 & (211)  & = & (112) & (\underline{\mathbf{0}11} 112) & \mathbf{6} \\
        5 & (201) & \oplus_3 & (221)  & = & (122) & (\underline{\mathbf{0}10} 122) & \mathbf{6} \\
        6 & (201) & \oplus_3 & (222)  & = & (120) & (\underline{\mathbf{1}20} 120) & \mathbf{6} \\
        7 & (201) & \oplus_3 & (022)  & = & (220) & (\underline{\mathbf{0}21} 220) & 7          \\
        8 & (201) & \oplus_3 & (002)  & = & (220) & (\underline{\mathbf{1}21} 200) & \mathbf{6} \\
    \end{array}
\end{equation*}
The underlined symbols represent the appended prefix, the bold underlined symbol is $u$, which is chosen such that $\beta_{6,3}$ is obtained whenever possible, and the bold weights indicate that balancing was achieved. Fig.~\ref{fig:ex2} presents the random walk graph for the weight of the overall sequence, $\bfc$, with the shaded area indicating the possible weights as a result of the flexibility in choosing $u$.
\end{example}

\subsubsection{Decoding}

The decoding consists of recovering the index $z$ from the Gray code prefix, $\bfg$, and finding $s$ and $p$ to reconstruct $\bfb_z$. The original sequence is then obtained as $\bfx = \bfy \ominus_q \bfb_z$, where $\ominus_q$ represents modulo $q$ subtraction.

As an example, Table~\ref{tab:a1} shows the decoding of every Gray code sequence into balancing sequences using the $(2,3)$-Gray code set. 
\begin{table}[!b]
	\centering
	\caption{Decoding of $(2,3)$-Gray codes for $3$-ary sequences of length 2}\label{tab:a1}
	\begin{tabular}{ccccc}
		\hline\hline
		Gray code ($\bfg$) & Sequence ($\bfd$) & $z$ & $s,p$ & $\bfb_z$ \\
		\hline
		$(00)$             & $(00)$            & 0   & $0,0$ & $(000)$  \\
		$(01)$             & $(01)$            & 1   & $0,1$ & $(100)$  \\
		$(02)$             & $(02)$            & 2   & $0,2$ & $(110)$  \\
		$(12)$             & $(10)$            & 3   & $1,0$ & $(111)$  \\
		$(11)$             & $(11)$            & 4   & $1,1$ & $(211)$  \\
		$(10)$             & $(12)$            & 5   & $1,2$ & $(221)$  \\
		$(20)$             & $(20)$            & 6   & $2,0$ & $(222)$  \\
		$(21)$             & $(21)$            & 7   & $2,1$ & $(022)$  \\
		$(22)$             & $(22)$            & 8   & $2,2$ & $(002)$  \\
		\hline\hline
	\end{tabular}	
\end{table}
\\
\begin{example}\label{ex3}
Consider the received ternary sequence $\bfc = (012012)$ of length $n=6$ (one of the balanced sequences from Example~\ref{ex2}). The $(2,3)$-Gray code prefixes were used in encoding the original sequence.

The first symbol in $\bfc$, $u=0$ is dropped, then the Gray code prefix is $\bfg=(12)$. This Gray code corresponds to $\bfd=(10)$ as presented in Table~\ref{tab:a1}. This implies that $z=3$, leading to $s=1$, $p=0$ and therefore $\bfb_3 = (111)$.
The original sequence is recovered as
\begin{equation*}
    \bfx = \bfy \ominus_q \bfb_z = (012) \ominus_3 (111) = (201).
\end{equation*}
Thus, the information sequence from Example~\ref{ex2} is recovered.
\end{example}

\section{Construction for $k \neq q^t$}\label{sec:4}

We will now generalize the technique described in the previous section to sequences of any length, i.e. $k \neq q^t$.  

The idea is to use a subset of the $(r',q)$-Gray code with an appropriate length to encode the $z$ indices that represent the $kq$ balancing sequences. Therefore, the cardinality of $(r',q)$-Gray code prefixes must be greater than that of the balancing sequences, i.e. $q^{r'} > kq$ or $r' > \log_q k + 1$.

However, the challenge is to find the appropriate subset of $(r',q)$-Gray code prefixes that can uniquely match the $kq$ balancing sequences, and still guarantee balancing when combined with $u$ and $\bfy$. 

\subsection{$(r',q)$-Gray code prefixes for $q$ odd}\label{sec:subset-q-odd}

\begin{figure}[!b]
	\centering
	\includegraphics[width=1\linewidth]{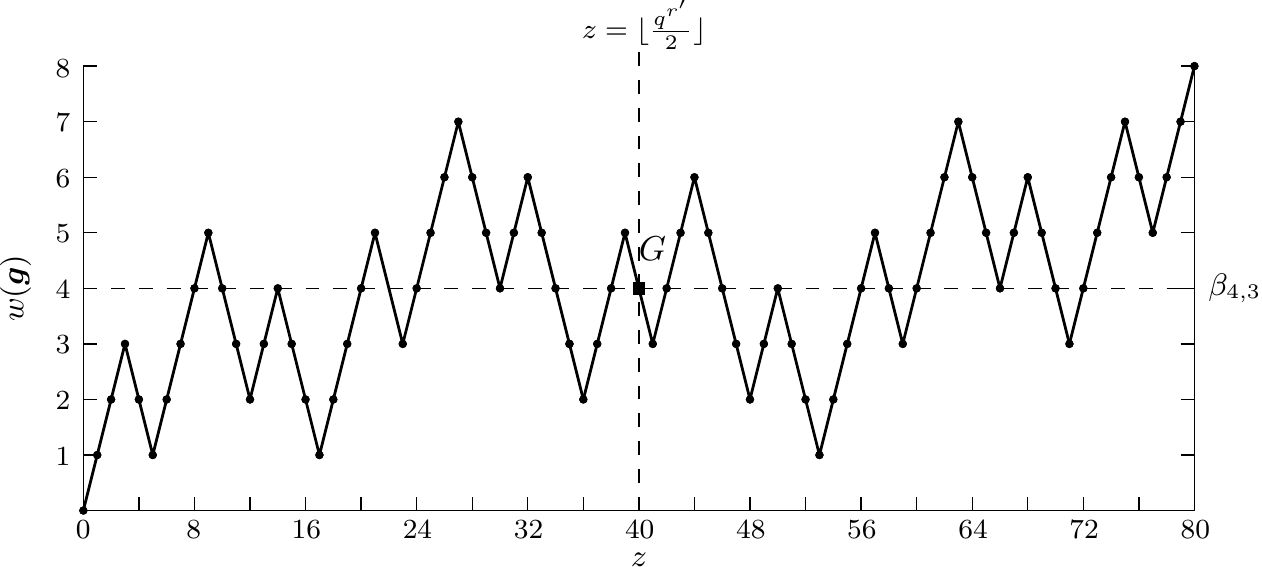}
	\caption{$(4,3)$-Gray code random walk graph}
	\label{fig:graycode1}
\end{figure}

When examining the random walk graph for Gray codes with $q$ odd, one notices that the random walk forms an odd function around a specific point. Fig.~\ref{fig:graycode1} presents the $(4,3)$-Gray code random walk graph, with $G$ being the intersection point between the horizontal line, $w(\bfg)=4$, and the vertical line, $z = 40$. The graph forms an odd function around this point $G$. In general, for $(r',q)$-Gray codes where $q$ is odd, the random walk of the Gray codes gives an odd function centered around $w(\bfg)=\beta_{r',q}$ and $z = \lfloor \frac{q^{r'}}{2} \rfloor$, where $\lfloor \cdot \rfloor$ represents the floor function.
\\
\begin{lemma}\label{lem:1}
The random walk graph of $(r',q)$-Gray codes where $q$ is odd forms an odd function around the point $G$.
\end{lemma}

It was proved in \cite{er1984} that any $(r',q)$-Gray code, where $q$ is odd, is reflected. That is, the random walk graph of the $(r',q)$-Gray code forms an odd function centered around the point $G$.

This implies that any subset of an $(r',q)$-Gray code around the center of its random walk graph, where the information sequence is such that $kq$ is odd (i.e. $k$ is odd), always has an average weight equal to $\beta_{r',q}$. As we need a unique subset of Gray code sequences for any case, we choose $kq$ elements from the ``middle'' values of $z \in [0,q^{r'}-1]$ and call it the $z$-centered subset. The index for this subset is denoted by $z'$, with $z' \in [z_1, z_2]$. When $kq$ is even (i.e. $k$ is even), it is not guaranteed that the subset of $(r',q)$-Gray codes' average weight around the center equals exactly $\beta_{r',q}$. However, it will be very close to it, with a rounded value that is equal to $\beta_{r',q}$. We formalize these observations in the subsequent lemma.
    
Let $\mathcal{G}$ denote the subset of $kq$ Gray code sequences that are used to encode the index $z'$, let $\overline{w}(\cdot)$ denote the average weight of a set of sequences and let $\lVert\cdot\rVert$ denote rounding to the nearest integer.\\

\begin{lemma}\label{lem:2}
For an $(r',q)$-Gray code subset, $\mathcal{G}$, where $q$ is odd and the $z'$-th codewords are chosen with $z' \in [z_1, z_2]$, the following holds:
\begin{itemize}
    \item if $k$ is odd with $z_1 = \lfloor \frac{q^{r'}}{2} \rfloor - \lfloor \frac{kq}{2} \rfloor$ and $z_2 = \lfloor \frac{q^{r'}}{2} \rfloor + \lfloor \frac{kq}{2} \rfloor$, then $\overline{w}(\mathcal{G}) = \beta_{r',q}$,
    \item if $k$ is even with $z_1 = \lfloor \frac{q^{r'}}{2} \rfloor - \frac{kq}{2}$ and $z_2 = \lfloor \frac{q^{r'}}{2} \rfloor + \frac{kq}{2}-1$, then $\lVert\overline{w}(\mathcal{G})\rVert = \beta_{r',q}$.
\end{itemize}
\end{lemma}

\begin{proof}
To simplify notation in this proof, we simply use $\beta$ to represent $\beta_{r',q}$ throughout.

If $k$ is odd, it follows directly from Lemma~\ref{lem:1} that choosing $kq$ sequences (where $kq$ is odd) from $z=\lfloor \frac{q^{r'}}{2} \rfloor - \lfloor \frac{kq}{2} \rfloor$ to $z=\lfloor \frac{q^{r'}}{2} \rfloor + \lfloor \frac{kq}{2} \rfloor$, centered around $z=\lfloor \frac{q^{r'}}{2} \rfloor$, will result in $\overline{w}(\mathcal{G}) = \beta$, since the random walk forms an odd function around this point.

In cases where $k$ is even, if $z_2$ was chosen as $\lfloor \frac{q^{r'}}{2} \rfloor + \frac{kq}{2}$, we would have exactly $\overline{w}(\mathcal{G}) = \beta$ (using the same reasoning as for the case where $k$ is odd), as we use $\frac{kq}{2}$ elements to the left of $\lfloor \frac{q^{r'}}{2} \rfloor$ and $\frac{kq}{2}$ elements to the right of it. However, this would mean that $kq+1$ elements are being used. Thus, $z_2 = \lfloor \frac{q^{r'}}{2} \rfloor + \frac{kq}{2}-1$ must be used. Let $\alpha$ be the weight of the $(z_2+1)$-th Gray code, then
\begin{equation*}
    \overline{w}(\mathcal{G}) = \frac{(kq+1)\beta - \alpha}{kq} = \beta + \frac{\beta-\alpha}{kq}.
\end{equation*}
The lowest possible value of $\alpha$ is $\alpha_{\min} = 0$, and its highest possible value is $\alpha_{\max} = k(q-1)$. Thus,
\begin{equation*}
    \beta + \frac{\beta-\alpha_{\max}}{kq} \leq \overline{w}(\mathcal{G}) \leq \beta + \frac{\beta-\alpha_{\min}}{kq}
\end{equation*}
and with some manipulations it can be shown that
\begin{equation*}
    \beta \left(1 - \frac{\frac{2k}{r'}-1}{kq}\right) \leq \overline{w}(\mathcal{G}) \leq \beta\left(1 + \frac{1}{kq}\right).
\end{equation*}
Finally, where $q$ is odd, we have $q \geq 3$, and rounding to the nearest integer results in $\lVert\overline{w}(\mathcal{G})\rVert = \beta$.
\end{proof}

\subsection{$(r',q)$-Gray code prefixes for $q$ even}\label{sec:subset-q-even}

\begin{figure}[h]
	\centering
	\includegraphics[width=1\linewidth]{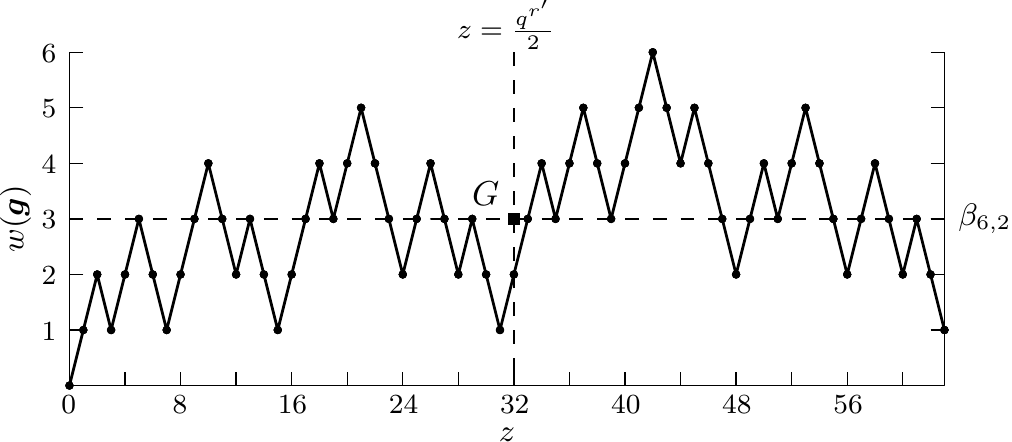}
	\caption{$(6,2)$-Gray code random walk graph}
	\label{fig:graycode2}
\end{figure}

For the encoding of sequences that make use of $(r',q)$-Gray code prefixes where $q$ is even, a different approach is followed. The subset of Gray code prefixes is obtained by placing a sliding window of length $kq$ over the random walk graph of the $(r',q)$-Gray code sequences, and shifting it until we obtain a subset with an average weight value of $\beta_{r',q}$. Fig.~\ref{fig:graycode2} shows the $(6,2)$-Gray code random walk graph.

However, this process does not always guarantee a subset of Gray code prefixes with an average weight value of exactly $\beta_{r',q}$. Since we have flexibility in choosing $u$, we can choose the average weight for the subset to be close to $\beta_{r',q}$, and adjust $u$ as necessary to obtain exact balancing.
\\
\begin{lemma}\label{lem:3}
An $(r', q)$-Gray code subset, $\mathcal{G}$, where $q$ is even, can be chosen such that $\lVert\overline{w}(\mathcal{G})\rVert = \beta_{r',q}$.
\end{lemma}
\begin{proof}
A similar reasoning as in the proof of Lemma~\ref{lem:2}, where a symbol with weight $\alpha$ is repeatedly removed from the set, can be used to find $\lVert\overline{w}(\mathcal{G})\rVert$.
\end{proof}

\subsection{Encoding}

Having presented all the required components, we now propose our encoding algorithm. The length of the required Gray code prefix is
\begin{equation}\label{eq1}
    r' = \lceil \log_qk \rceil + 1,
\end{equation}
where $\lceil \cdot \rceil$ represents the ceiling function.

The cardinality of $(r',q)$-Gray codes equals $q^{r'}$. This implies that $q^{r'-1} < kq < q^{r'}$. The encoding will make use of a subset of $kq$ Gray code sequences from the $q^{r'}$ available ones.\\

\begin{theorem}
Any $q$-ary sequence can be balanced by adding (modulo $q$) an appropriate balancing sequence, $\bfb_z$, and prefixing a redundant symbol, $u$, with a Gray code sequence, $\bfg$, taken from the subset of $(r',q)$-Gray code prefixes.
\end{theorem}

\begin{proof}
Let $\mathcal{U}$ denote the set of possible symbols for $u$, i.e. $\mathcal{U} = \{0,1,\ldots,q-1\}$, let $\mathcal{G}$ denote the subset of Gray code sequences, and let $\mathcal{Y}$ denote the set of $kq$ output sequences after the $kq$ balancing sequences are added to the information sequence.

It is easy to see that
\begin{equation*}
    \overline{w}(\mathcal{U}) = \frac{(q-1)}{2}.
\end{equation*}
From Lemmas~\ref{lem:2} and \ref{lem:3}, the subset of $(r',q)$-Gray code prefixes that corresponds to the $kq$ balancing sequences is chosen such that	
\begin{equation*}
    \lVert\overline{w}(\mathcal{G})\rVert = \frac{r'(q-1)}{2} = \beta_{r',q}.
\end{equation*}
It was proved in \cite{swart2009} that the average weight of the $kq$ sequences, $\bfy = \bfx \oplus_q \bfb_z$, is such that
\begin{equation*}
    \overline{w}(\mathcal{Y}) = \frac{k(q-1)}{2} = \beta_{k,q}.
\end{equation*}
	
By considering $\bfc = (u|\bfg|\bfy)$, with length $n = k+r'+1$, as the overall sequence to be transmitted, it follows that:
\begin{align*}
    \overline{w}(\mathcal{U}) + \lVert\overline{w}(\mathcal{G})\rVert + \overline{w}(\mathcal{Y}) &= \frac{(q-1)}{2} + \frac{r'(q-1)}{2} + \frac{k(q-1)}{2} \\
                                                   &= \frac{(k+r'+1)(q-1)}{2} \\
                                                   &= \beta_{n,q}.
\end{align*}

This implies that there is at least one $\bfc$ for which $w(\bfc) \leq \beta_{n,q}$ and at least one other $\bfc$ for which $w(\bfc) \geq \beta_{n,q}$. Taking the random walk's increases into account, as well as the flexibility in choosing $u$, we can conclude that there is at least one $\bfc$ such that $w(\bfc) = \beta_{n,q}$.
\end{proof}   

The encoding algorithm consists of the following steps:
\begin{enumerate}
    \item Obtain the correct Gray code length $r'$ by using \eqref{eq1}. Then find the corresponding subset of $(r',q)$-Gray code prefixes, $z' \in [z_1, z_2]$, using the methods discussed in Section~\ref{sec:subset-q-odd} where $q$ is odd and in Section~\ref{sec:subset-q-even} where $q$ is even.
    \item Incrementing through $z$, determine the balancing sequences, $\bfb_z$, and add them to the information sequence $\bfx$ to obtain outputs $\bfy$.
    \item For each increment of $z$, append every $\bfy$ with the corresponding Gray code prefix $\bfg$ following the lexicographic order, with $\bfg$ obtained from the $q$-ary representations of the $z'$ indices.
    \item Finally, set $u = \beta_{n,q}-w(\bfy)-w(\bfg)$ if $u \in \{0, 1, \ldots, q-1\}$, otherwise set $u=0$.
\end{enumerate}
We illustrate the encoding algorithm with the following two examples, one for an odd value of $q$ and the other for an even value of $q$.\\

\begin{example}\label{ex4}
Consider encoding the ternary sequence, $(21120)$, of length 5. Since $r' = \lceil \log_3 5 \rceil + 1 = 3$, we require $(3,3)$-Gray code prefixes to encode the $z'$ indices. The overall sequence length is $n = k+r'+1 = 9$, and the balancing value is $\beta_{9,3} = 9$. The cardinality of the $(3,3)$-Gray code is 27 and the required $z$-centered subset of prefixes containing $kq=15$ elements is such that $z' \in [5,19]$.

The following process shows the possible sequences obtained. Again the underlined symbols represent the appended prefix, the bold underlined symbol is $u$, and the bold weights indicate balancing.
\begin{equation*}
	\begin{array}{c@{\;}ccc@{\;}c}
        z  & z' & \bfx \oplus_q \bfb_z =\bfy & \bfc&w(\bfc)\\
        \hline
        0  & 5  & (21120)  \oplus_3  (00000)  =  (21120) & (\underline{\mathbf{2}010}21120) & \mathbf{9} \\
        1  & 6  & (21120)  \oplus_3  (10000) =  (01120) & (\underline{\mathbf{0}020}01120) & 6          \\
        2  & 7  & (21120)  \oplus_3  (11000)  =  (02120) & (\underline{\mathbf{1}021}02120) & \mathbf{9} \\
        3  & 8  & (21120)  \oplus_3 (11100)  =  (02220) & (\underline{\mathbf{0}022}02220) & 10         \\
        4  & 9  & (21120)  \oplus_3  (11110)  =  (02200) & (\underline{\mathbf{0}122}02200) & \mathbf{9} \\
        5  & 10 & (21120)  \oplus_3  (11111)  =  (02201) & (\underline{\mathbf{0}121}02201) & \mathbf{9} \\
        6  & 11 & (21120)  \oplus_3  (21111)  =  (12201) & (\underline{\mathbf{0}120}12201) & \mathbf{9} \\
        7  & 12 & (21120)  \oplus_3  (22111)  =  (10201) & (\underline{\mathbf{0}110}10201) & 6          \\
        8  & 13 & (21120)  \oplus_3  (22211)  =  (10001) & (\underline{\mathbf{0}111}10001) & 5          \\
        9  & 14 & (21120)  \oplus_3  (22221)  =  (10011) & (\underline{\mathbf{2}112}10011) & \mathbf{9} \\
        10 & 15 & (21120) \oplus_3  (22222)  =  (10012) & (\underline{\mathbf{2}102}10012) & \mathbf{9} \\
        11 & 16 & (21120)  \oplus_3  (02222)  =  (20012) & (\underline{\mathbf{2}101}20012) & \mathbf{9} \\
        12 & 17 & (21120)  \oplus_3  (00222)  =  (21012) & (\underline{\mathbf{2}100}21012) & \mathbf{9} \\
        13 & 18 & (21120)  \oplus_3  (00022)  =  (21112) & (\underline{\mathbf{0}200}21112) & \mathbf{9} \\
        14 & 19 & (21120)  \oplus_3  (00002)  =  (21122) & (\underline{\mathbf{0}201}21122) & 11         \\
    \end{array}
\end{equation*}
\end{example}

\begin{example}\label{ex5}
Consider encoding the 4-ary sequence, (312), of length 3. As before, $r' = \lceil \log_4 3 \rceil + 1 = 2$, requiring $(2,4)$-Gray code prefixes to be used. The overall sequence length is $n = 6$, and the balancing value is $\beta_{6,4} = 9$. The cardinality of the $(2,4)$-Gray code equals 16. The $z'$-subset is found by employing a sliding window of length $kq = 12$ over the random walk graph of the $(2,4)$-Gray code prefixes, shown in Fig.~\ref{fig:graycode3}. A suitable subset is found where $z_1 = 1$ and $z_2 = 12$, with an average weight value of 3, which equals $\beta_{2,4} = 3$.
\begin{figure}[!b]
    \centering
    \includegraphics[width=1\linewidth]{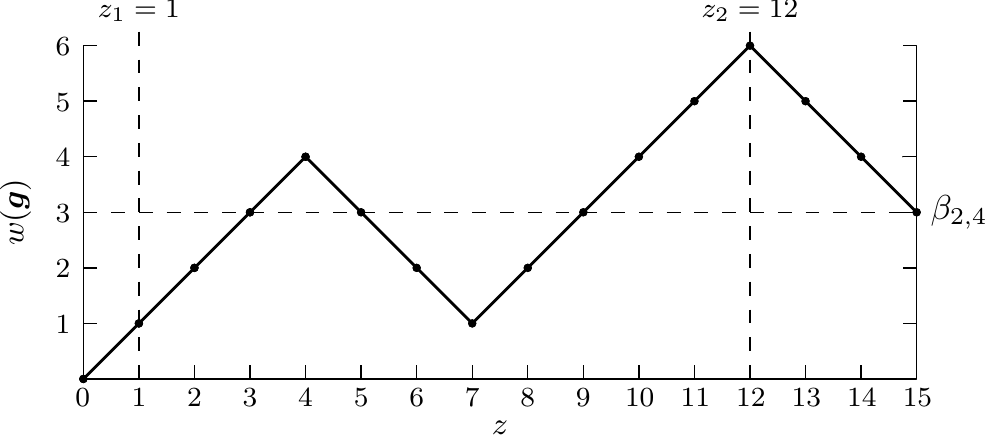}
    \caption{$(2,4)$-Gray code random walk graph with chosen subset}
    \label{fig:graycode3}
\end{figure}

The encoding process for the 4-ary sequence is shown next.
\begin{equation*}
	\begin{array}{c@{\quad}c@{\quad\quad}c@{\;}c@{\;}c@{\;}c@{\;}c@{\quad\quad}c@{\quad\quad}c}
        z  & z' & \bfx  & \oplus_q & \bfb_z & = & \bfy  & \bfc                          & w(\bfc)    \\
        \hline
        0  & 1  & (312) & \oplus_4 & (000)  & = & (312) & (\underline{\mathbf{2}01}312) & \mathbf{9} \\
        1  & 2  & (312) & \oplus_4 & (100)  & = & (012) & (\underline{\mathbf{0}02}012) & 5          \\
        2  & 3  & (312) & \oplus_4 & (110)  & = & (022) & (\underline{\mathbf{2}03}022) & \mathbf{9} \\
        3  & 4  & (312) & \oplus_4 & (111)  & = & (023) & (\underline{\mathbf{0}13}023) & \mathbf{9} \\
        4  & 5  & (312) & \oplus_4 & (211)  & = & (123) & (\underline{\mathbf{0}12}123) & \mathbf{9} \\
        5  & 6  & (312) & \oplus_4 & (221)  & = & (133) & (\underline{\mathbf{0}11}133) & \mathbf{9} \\
        6  & 7  & (312) & \oplus_4 & (222)  & = & (130) & (\underline{\mathbf{0}10}130) & 5          \\
        7  & 8  & (312) & \oplus_4 & (322)  & = & (230) & (\underline{\mathbf{2}20}230) & \mathbf{9} \\
        8  & 9  & (312) & \oplus_4 & (332)  & = & (200) & (\underline{\mathbf{0}21}200) & 5          \\
        9  & 10 & (312) & \oplus_4 & (333)  & = & (201) & (\underline{\mathbf{2}22}201) & \mathbf{9} \\
        10 & 11 & (312) & \oplus_4 & (033)  & = & (301) & (\underline{\mathbf{0}23}301) & \mathbf{9} \\
        11 & 12 & (312) & \oplus_4 & (003)  & = & (311) & (\underline{\mathbf{0}33}311) & 11         \\
    \end{array}		
\end{equation*}
\end{example}

\subsection{Decoding}

\begin{figure}[!t]
	\centering
	\includegraphics{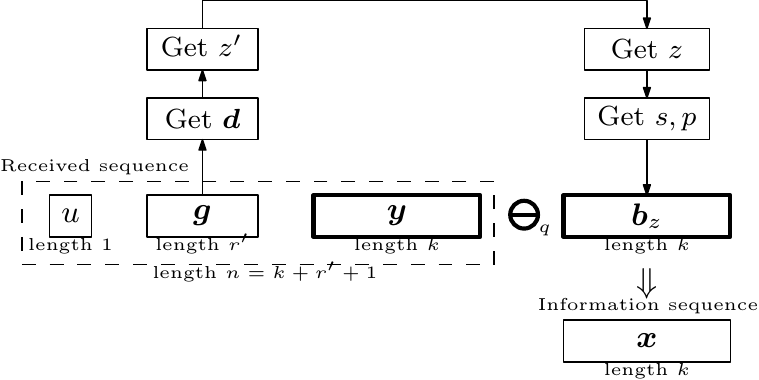}
	\caption{Decoding process for any $q$-ary information sequence} 
	\label{fig:dec2}
\end{figure}

Fig.~\ref{fig:dec2} presents the decoding process of our proposed scheme, for any $q$-ary information sequence. The decoding algorithm consists of the following steps:
\begin{enumerate}
	\item The redundant symbol $u$ is dropped, then the following $r'$ symbols are extracted as the Gray code prefix, $\bfg$, converted to $\bfd$ and used to find $z'$.
	\item From $z'$, the corresponding $z$ index is computed as $z=z'-z_1$.
	\item $z$ is used to find the parameters $s$ and $p$, then $\bfb_z$ is derived.
	\item Finally, the original sequence is recovered through $\bfx = \bfy \ominus_q \bfb_z$.
\end{enumerate}

\begin{example}\label{ex6}
Consider the decoding of the sequence, $(\underline{\mathbf{2}100}121200)$ (the underlined symbols are the prefix and the bold underlined symbol is $u$), where $n=10$ and $q=3$, that was encoded using $(3,3)$-Gray code prefixes.

The first symbol $u=2$ is dropped, then the Gray code prefix is extracted as $(100)$, which corresponds to $z'=17$, and the $z'$-subset of $(3,3)$-Gray code prefixes is $z' \in [4,21]$, thus $z=13$. This can be seen from Table~\ref{tab:a2}, where the decoding of all $(3,3)$-Gray codes is shown.

This implies that $s=2$ and $p=1$, resulting in $\bfb_{13}=(022222)$. Finally, the original information sequence is extracted as $\bfx = \bfy \ominus_q \bfb_z = (121200) \ominus_3 (022222) = (102011)$.
\begin{table}[!b]
	\centering
	\caption{Decoding of $(3,3)$-Gray codes for ternary sequences with $k=6$ and $z' \in [4,21]$}\label{tab:a2}
	\begin{tabular}{cccccc}
		\hline\hline
		Gray code ($\bfg$) & Sequence ($\bfd$) & $z'$ & $z$     & $s,p$ & $\bfb_z$   \\
		\hline
		$(000)$            & $(000)$               & 0    & --- & ---   & ---        \\
		$(001)$            & $(001)$               & 1    & --- & ---   & ---        \\
		$(002)$            & $(002)$               & 2    & --- & ---   & ---        \\
		$(012)$            & $(010)$               & 3    & --- & ---   & ---        \\
		\hline
		$(011)$            & $(011)$               & 4    & 0   & $0,0$ & $(000000)$ \\
		$(010)$            & $(012)$               & 5    & 1   & $0,1$ & $(100000)$ \\
		$(020)$            & $(020)$               & 6    & 2   & $0,2$ & $(110000)$ \\
		$(021)$            & $(021)$               & 7    & 3   & $0,3$ & $(111000)$ \\
		$(022)$            & $(022)$               & 8    & 4   & $0,4$ & $(111100)$ \\
		$(122)$            & $(100)$               & 9    & 5   & $0,5$ & $(111110)$ \\
		$(121)$            & $(101)$               & 10   & 6   & $1,0$ & $(111111)$ \\
		$(120)$            & $(102)$               & 11   & 7   & $1,1$ & $(211111)$ \\
		$(110)$            & $(110)$               & 12   & 8   & $1,2$ & $(221111)$ \\
		$(111)$            & $(111)$               & 13   & 9   & $1,3$ & $(222111)$ \\
		$(112)$            & $(112)$               & 14   & 10  & $1,4$ & $(222211)$ \\
		$(102)$            & $(120)$               & 15   & 11  & $1,5$ & $(222221)$ \\
		$(101)$            & $(121)$               & 16   & 12  & $2,0$ & $(222222)$ \\
		$(100)$            & $(122)$               & 17   & 13  & $2,1$ & $(022222)$ \\
		$(200)$            & $(200)$               & 18   & 14  & $2,2$ & $(002222)$ \\
		$(201)$            & $(201)$               & 19   & 15  & $2,3$ & $(000222)$ \\
		$(202)$            & $(202)$               & 20   & 16  & $2,4$ & $(000022)$ \\
		$(212)$            & $(210)$               & 21   & 17  & $2,5$ & $(000002)$ \\
		\hline
		$(211)$            & $(211)$               & 22   & --- & ---   & ---        \\
		$(210)$            & $(212)$               & 23   & --- & ---   & ---        \\
		$(220)$            & $(220)$               & 24   & --- & ---   & ---        \\
		$(221)$            & $(221)$               & 25   & --- & ---   & ---        \\
		$(222)$            & $(222)$               & 26   & --- & ---   & ---        \\
		\hline\hline
	\end{tabular}	
\end{table}
\end{example}

\section{Redundancy and Complexity}\label{sec:5}

In this section we compare the redundancy and complexity of our proposed scheme with some existing ones.

\subsection{Redundancy}\label{sec:5.1}

Let $\mathcal{F}_q^k$ denote the cardinality of the full set of balanced $q$-ary sequences of length $k$. According to \cite{star1975},
\begin{equation*}
    \mathcal{F}_q^k = q^{k} \sqrt{\frac{6}{\pi r(q^2-1)}} \left(1+\mathcal{O}\left(\frac{1}{k}\right)\right).
\end{equation*} 

The information sequence length, $k$, in terms of the redundancy, $r$, for the construction in \cite{swart2009} is
\begin{equation}\label{eq_c1}
    k \leq \frac{\mathcal{F}_q^r}{q} \approx q^{r-1} \sqrt{\frac{6}{\pi r(q^2-1)}}.
\end{equation}

In \cite{capocelli1991}, two schemes are presented for $k$ information symbols, where one satisfies the bound
\begin{equation}\label{eq_c3}
	k \leq \frac{q^r-1}{q-1},
\end{equation}
and the other one satisfies 
\begin{equation}\label{eq_c2}
	k \leq 2\frac{q^r-1}{q-1}-r.
\end{equation}

The construction in \cite{tallini1999} presents the information sequence length in terms of the redundancy as
\begin{equation*}
    k \leq \frac{1}{1-2\gamma}\frac{q^r-1}{q-1}-a_1(q, \gamma)r-a_2(q, \gamma),
\end{equation*}
with $\gamma \in [0,\frac{1}{2})$, where $a_1$ and $a_2$ are scalars depending on $q$ and $\gamma$. If the compression aspect is ignored, the information sequence length is the same as in \eqref{eq_c2}.

The prefixless scheme presented in \cite{swart2013} has information sequence length that satisfies
\begin{equation}\label{eq_c4}
    k \leq q^{r-1} - r.
\end{equation}   

Two constructions with parallel decoding are presented in \cite{pelusi2015}. The first construction, where the prefixes are also balanced as in \cite{swart2009}, has its information length as a function of $r$ as
\begin{equation}\label{eq_c5}
    k \leq \frac{\mathcal{F}_q^r - \{ q \bmod 2 + [(q-1)k] \bmod 2\}}{q-1}.
\end{equation}   
The second construction, where the prefixes need not be balanced, is a refinement of the first and has an information length the same as \eqref{eq_c3}.

As presented in Section~\ref{sec:4}, the redundancy of our new construction is given by $r = \lceil \log_q k \rceil + 2$. Therefore, the information sequence length in terms of redundancy is
\begin{equation}
    k = q^{r-2}.
\end{equation}

\begin{figure}[!t]
	\centering
	\includegraphics[width=1\linewidth]{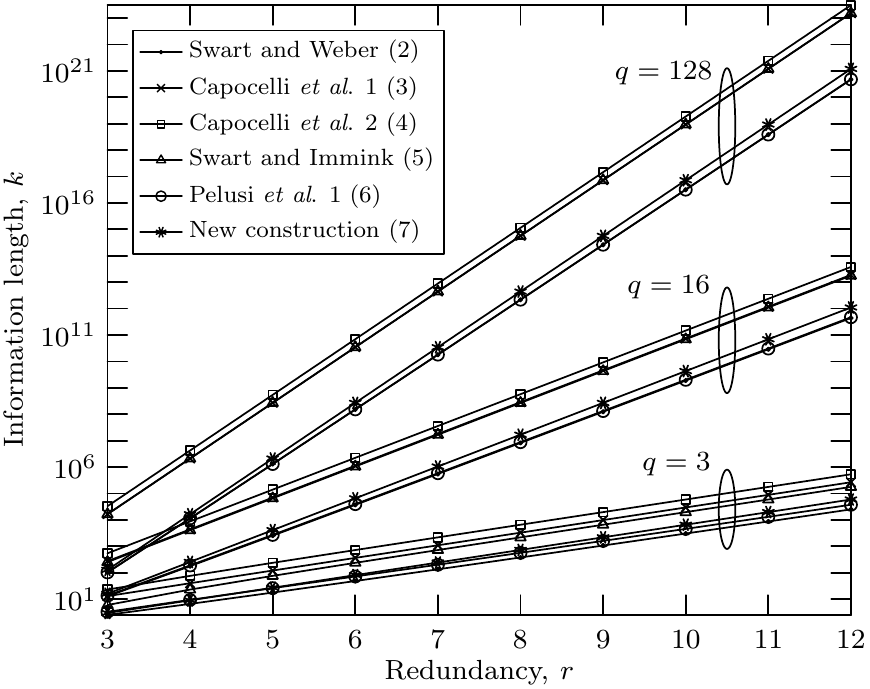}
	\caption{Comparison of information sequence length vs. redundancy for various schemes}
	\label{fig:redundancies}
\end{figure}

Fig.~\ref{fig:redundancies} presents a comparison of the information length, $k$, versus the redundancy, $r$, for various constructions as discussed above. For all $q$, our construction is only comparable to the information lengths from \eqref{eq_c1} and \eqref{eq_c5}, although it does slightly improve on both.

However, the trade-off is that as the redundancy becomes greater, the complexity of our scheme tends to remain constant, as we see in the next section.

\subsection{Complexity}\label{sec:5.2}

We estimate the complexity of our proposed scheme and compare it to that of existing algorithms.

The techniques in \cite{capocelli1991} and \cite{tallini1999} both require $\mathcal{O}(qk\log_qk)$ digit operations for the encoding and decoding. The method from \cite{swart2009} takes $\mathcal{O}(qk\log_qk)$ digit operations for the encoding and $\mathcal{O}(1)$ digit operations for the decoding. A refined design of the parallel decoding method is presented in \cite{pelusi2015}, where the complexity equals $\mathcal{O}(k\sqrt{\log_qk})$ in the encoding case and $\mathcal{O}(1)$ digit operations in the decoding process.

The following pseudo code presents the steps of our encoding method:
\begin{verbatim}
Input: Information sequence, x of length k.
Output: Encoded sequence, y of length n=k+r.

for i=0:kq;
   for j=z1:z2;
      y(i) = [u | g(j) |  x + b(s,p)(i)];
        If (w(y(i))==beta)
          // Testing for balanced sequence.
          exit();
          //Terminate the program.
   end;
end;
\end{verbatim}

In the above code, $i$ is the iterator through the $kq$ output sequences and also through the balancing sequences, while $j$ is the iterator through the subset of Gray code sequences, ranging from $z_1 =  \lfloor \frac{q^{r'}}{2} \rfloor - \lfloor \frac{kq}{2} \rfloor$ to $z_2= z_1 + kq-1$. The symbol `$\mid$' denotes the concatenation.

Our encoding scheme is based on the construction in \cite{swart2009} that has an encoding complexity of $\mathcal{O}(qk\log_qk)$, and it takes $\mathcal{O}(\log_qk)$ to encode Gray code prefixes as presented in \cite{guan1998}. Therefore the encoding of our algorithm requires $\mathcal{O}(qk\log_qk)$ digit operations.

The decoding process consists of very simple steps: the recovery of the index $z'$ from the Gray code requires $\mathcal{O}(\log_q k)$ digit operations \cite{guan1998}. After obtaining the index $z'$ from the Gray code prefix, the balancing sequence $\bfb_z$ is found and then the original information sequence is recovered through the operation, $\bfy = \bfx \ominus_q \bfb_z$, which can be performed in parallel, resulting in a complexity of $\mathcal{O}(1)$. Therefore the overall complexity for the decoding is $\mathcal{O}(\log_q k)$ digit operations.

Table~\ref{tab:cmpl} summarizes the complexities for various constructions, where the orders of digit operations it takes to complete the encoding/decoding are compared.
\begin{table}[!t]
	\centering
	\caption{Complexities of various schemes (orders are in digit operations)}\label{tab:cmpl}
    \renewcommand\arraystretch{1.5}
	\begin{tabular}{ccc}
		\hline\hline
		Algorithm            & Encoding order                 & Decoding order           \\
		\hline
		\cite{tallini1999}   & $\mathcal{O}(qk\log_qk)$       & $\mathcal{O}(qk\log_qk)$ \\
		\cite{capocelli1991} & $\mathcal{O}(qk\log_qk)$       & $\mathcal{O}(qk\log_qk)$ \\
		\cite{swart2009}     & $\mathcal{O}(qk\log_qk)$       & $\mathcal{O}(1)$         \\
		\cite{pelusi2015}    & $\mathcal{O}(k\sqrt{\log_qk})$ & $\mathcal{O}(1)$         \\
		Our scheme           & $\mathcal{O}(qk\log_qk)$       & $\mathcal{O}(\log_qk)$   \\
		\hline\hline
	\end{tabular}	
\end{table}

\section{Conclusion}\label{sec:6}

An efficient construction has been proposed for balancing non-binary information sequences. By making use of Gray codes for the prefix, no lookup tables are used, only linear operations are needed for the balancing and the Gray code implementation. The encoding scheme has a complexity of $\mathcal{O}(qk\log_qk)$ digit operations. For the decoding process, once the Gray code prefix is decoded using $\mathcal{O}(\log_qk)$ digit operations, the balancing sequence is determined and the rest of the decoding process is performed in parallel. This makes the decoding fast and efficient.

Possible future research directions include finding a mathematical procedure to determine the subset of Gray code sequences for $q$ even, given that it was found manually, by using a sliding window over the random walk graph. Practically, the redundant symbol $u$ only needs to take on values of zero (when the random walk falls on the balancing value) or one (when the random walk falls just below the balancing value). Thus, unnecessary redundancy is contained in $u$, especially for large values of $q$. However, the flexibility over $u$ increases the occurrences of balanced sequences. These additional balanced outputs could potentially be used to send auxiliary data that could reduce the redundancy. This property was proved for the binary case \cite{weber2010}. Additionally, given that the random walk graph passes through other weights in the region of the balancing value, the scheme can be extended to the construction of constant weight sequences with arbitrary weights.


\end{document}